\documentclass{iaus}
\usepackage{graphicx}

\title[Structure of the X-ray Jet in 4C19.44] %% give here short title %%
{Detailed Structure of the X-ray Jet in 4C 19.44 (=PKS1354+195)}

\author[D. A. Schwartz et al.]   %% give here short author list %%
{D. A. Schwartz$^1$, D. E. Harris$^1$, H. Landt$^1$, 
A. Siemiginowska$^1$, E. S. Perlman$^2$, C. C. Cheung$^3$, J. M. 
Gelbord$^4$, D. M. Worrall$^5$, M. Birkinshaw$^5$, S. G. Jorstad$^6$, 
A. P. Marscher$^6$ \and L. Stawarz$^7$ 
 }

\affiliation{$^1$Smithsonian Astrophysical Observatory, Cambridge MA
02138, USA \break email: das@cfa.harvard.edu \break
$^2$U. of MD. Baltimore County, Baltimore MD 21250, USA\break
$^3$NRAO/Stanford University, Palo Alto CA 94305, USA \break
$^4$Massachusetts Institute of Technology, Cambridge MA 02139, USA
\break
$^5$Physics Department, University of Bristol, Bristol BS8 1TL UK \break
$^6$Institute for Astrophysical Research, Boston University, Boston MA
02215, USA \break
$^7$Kavli Institute for Particle Astrophysics and Cosmology, Stanford
University, Stanford CA 94305, USA }\break

\pubyear{2006}
\volume{238}  %% insert here IAU Symposium No.
\pagerange{1--2}
\date{?? and in revised form ??}
\jname{Proceedings Title IAU Symposium}
\editors{A.C. Editor, B.D. Editor \& C.E. Editor, eds.}
\begin{document}

\maketitle

\begin{abstract}
We investigate the variations of the magnetic field, Doppler factor,
and relativistic particle density along the jet of a quasar at
z=0.72. We chose 4C 19.44 for this study because of its length and
straight morphology. The 18 arcsec length of the jet provides many
independent resolution elements in the Chandra X-ray image. The
straightness suggests that geometry factors, although uncertain, are
almost constant along the jet. We assume the X-ray emission is from inverse
Compton scattering of the cosmic microwave background. With the aid of
assumptions about jet alignment, equipartition between magnetic-field
and relativistic-particle energy, and filling factors, we find that
the jet is in bulk relativistic motion with a Doppler factor $\approx$
6 at an angle no more than 10$^\circ$ to the line of sight over
deprojected distances $\approx$ 150--600 kpc from the quasar, and with
a magnetic field $\approx$10 $\mu$Gauss.

\end{abstract}

\firstsection % if your document starts with a section,
\section{Analysis of the jet}\label{analysis}

\begin{figure}
\scalebox{0.97}{
\includegraphics{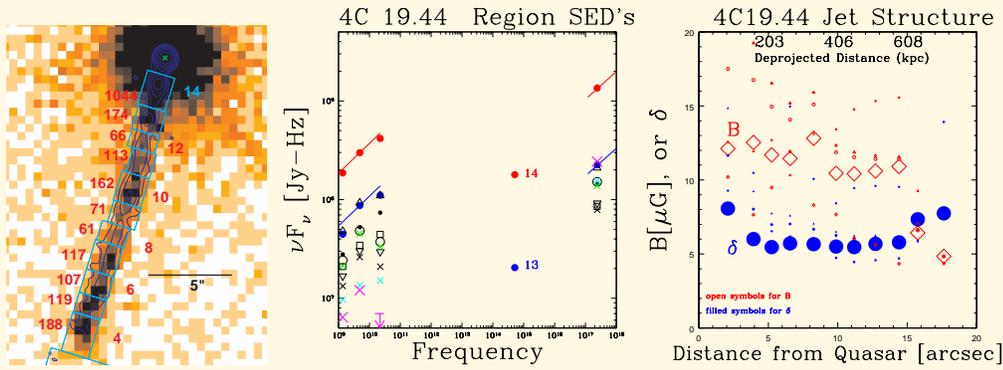} }
  \caption{Left panel shows the \emph{Chandra} 0.5 to 7 keV band X-ray
image of the 18 arcsec long jet. The boxes numbered 4 through 14 are an
arbitrary division of the jet into distinct regions for independent
analysis. Numbers to the left of each box show how many X-ray photons
were detected in that region. The contours are of the 4.8 GHz
emission. The middle panel shows the spectral energy distributions for
these 11 regions. Optical emission is detected only from the regions 14 and 13
(knots A and B of Sambruna et al. 2004). The summed X-ray spectrum
of all the regions is consistent with $f_{\nu}\propto \nu^{-0.7}$, and
the corresponding $\nu f_{\nu}$ is shown as the solid 
lines. The optical detection from region 13 can serve as an upper
limit to any region, and along with the radio indices prohibits a
simple synchrotron spectrum being 
extrapolated from the radio to X-ray region. The right hand panel
shows the results for the magnetic field (large  diamonds) and Doppler factor
(filled circles) as a function of distance along the jet. Alternate
assumptions can lead 
to systematic changes in these parameters  as follows:
small crosses use the best-fit power law radio spectral index;
small triangles and circles assume conical jets starting at the quasar
and 1 kpc from the quasar, respectively, reaching 1'' diameter at the
end of the jet in each case;
small diamonds use the 1.4 to 4.86 GHz spectral index instead of the
assumed slope of 0.7.
}\label{fig:regions} 
\end{figure}

We present  preliminary results from a 200 ks observation of the 
quasar 4C 19.44 (=PKS 1354+135), using \emph{Chandra}
observation identifications (
OBSID's) 6903, 6904, 7302, 7303. We analyze the jet  assuming X-ray
production by inverse Compton scattering on the cosmic microwave
background . This is motivated in the middle panel of
Figure~\ref{fig:regions} since the optical 
flux for two knots (from \cite{Sambruna04}) do not allow an
extrapolation of the 
synchrotron radio spectrum to the X-ray region. We follow Tavecchio et
al. (2000), and Celotti et al. (2001) to calculate the rest frame
magnetic field B and the Doppler factor $\delta=[\Gamma(1-\beta
\cos(\theta))]^{-1}$. The formalism requires many 
assumptions on the physics and on unknown values of
parameters. Consistent with our previous work (Schwartz et al. 2006)
we assume a minimum total energy in magnetic field and relativistic
particles, filling factor =1, an electron spectral index m=2.4 with
cutoffs at low and high energies tuned to give radio emission from
10$^6$ Hz to 10$^{12}$ Hz, equal energy in electrons and protons, and that
the bulk Lorentz factor $\Gamma=\delta$ so that the angle to the line
of sight takes on the maximum value, $\approx 1/\delta$, for a given
$\delta$. We take the volume of regions 4 through 14 to be cylinders
of the (resolved) lengths shown in the left panel of
Figure~\ref{fig:regions}, and (unresolved) radii assumed to be
0{\mbox{$.\!\!^{\prime\prime}$}}5 = 3.62 kpc at the redshift z=0.72 of
this quasar.

\section{Structure of the Jet}\label{structure}

The results for B and $\delta$ are presented in the rightmost panel of
the figure. With the above assumptions the jet shows similar structure
along its length. We give the deprojected distance on the top axis by
assuming $\delta =5.5$.  We note that if the radii were all 1 kpc, the
values for B and $\delta$ would be a factor of 1.44 higher.  The
systematic effects of our assumptions allow alternate profiles, but
the numerical values are constrained within a factor roughly 2. In
particular, the increase in $\delta$ at the end of the jet would be
peculiar in the IC/CMB scenario.  For the baseline values of B and
$\delta$, we can derive the kinetic power, the minimum electron
Lorentz factor, and the relativistic electron density as a function of
distance along the jet. For deprojected distances 200 to 600 kpc from
the quasar, regions 13 to 6, we require a cutoff to the
electron spectrum at energies below $80 mc^2$ and we estimate a kinetic
power of approximately (1--2) $\times$ 10$^{46}$ erg s$^{-1}$, and a
total relativistic electron density of roughly (2--3) $\times$
10$^{-8}$ cm$^{-3}$.

\begin{acknowledgments}
We acknowledge  support of NASA contract NAS8 39073 to SAO, 
CXC grants GO5-6116B,  GO6-7111A, GO6-7111B, and HST grant HST-GO-10762.01-A.
\end{acknowledgments}


\begin{thebibliography}{}

\bibitem[Celotti \etal\ (2001)]{Celotti01} Celotti, A., Ghisellini,
G., \& Chiaberge, M. 2001, \textit{MNRAS} 321, L1, astro-ph/0008021

\bibitem[Sambruna \etal\ (2004)]{Sambruna04}Sambruna, R. M., et
al. 2004 \textit{ApJ} 608, 698, astro-ph/0401475

\bibitem[Schwartz \etal\ (2006)]{Schwartz06} Schwartz, D. A., et
al. 2006, \textit{ApJ} 640, 592, astro-ph/0601632 

\bibitem[Tavecchio \etal\ (2000)]{Tavecchio00}
Tavecchio, F., Maraschi, L., Sambruna, R. M., Urry, C. M. 2000,
\textit{ApJ} 544, L23, astro-ph/0007441


\end{thebibliography}
\end{document}